\documentclass[journal=inoraj,manuscript=article]{achemso}

\usepackage{colortbl}
\usepackage{float}
\usepackage{dcolumn}
\usepackage{bm}
\usepackage{color}
\usepackage{mathtools}
\usepackage{subfig}
\usepackage{tabularx} 
\usepackage{graphicx}
\usepackage[hidelinks]{hyperref}

\usepackage{blindtext}
\usepackage{xcolor}
\usepackage[normalem]{ulem}               
\newcommand\blueout{\bgroup\markoverwith
{\textcolor{blue}{\rule[0.5ex]{2pt}{0.8pt}}}\ULon}

\usepackage{xr}
\usepackage{soul}
\title{Core excitations of uranyl in Cs$_{2}$UO$_{2}$Cl$_{4}$ from relativistic embedded damped-response time-dependent density functional theory calculations}

\author{Wilken Aldair Misael}
\author{Andr\'{e} Severo Pereira Gomes}
\email{andre.gomes@univ-lille.fr}
\affiliation{Univ. Lille, CNRS, UMR 8523-PhLAM-Physique des Lasers Atomes et Mol\'{e}cules, F-59000 Lille, France}

\begin{document}


\begin{abstract}
X-ray spectroscopies, by their high selectivity and sensitivity to the chemical environment around the atoms probed, provide significant insight into the electronic structure of molecules and materials. Interpreting experimental results requires reliable theoretical models, accounting for environment, relativistic, electron correlation, and orbital relaxation effects in a balanced manner. 
In this work, we present a protocol for the simulation of core excited spectra with damped response time-dependent density functional theory based on the Dirac-Coulomb Hamiltonian (4c-DR-TD-DFT), in which environment effects are accounted for through the frozen density embedding (FDE) method. 
We showcase this approach for the uranium M$_4$-, L$_3$-edge and oxygen K-edge of uranyl tetrachloride (UO$_2$Cl$_4^{2-}$) unit as found in a host Cs$_{2}$UO$_{2}$Cl$_{4}$ crystal. 
We have found that the 4c-DR-TD-DFT simulations yield excitation spectra that very closely match the experiment for the uranium M$_4$- and oxygen K-edges, with good agreement for the broad experimental spectra for the L$_3$-edge. By decomposing the complex polarizability in terms of its components we have been able to correlate our results with angle-resolved spectra. 
We have observed that for all edges, but in particular the uranium M$_4$-edge, an embedded model in which the chloride ligands are replaced by an embedding potential, reproduces rather well the spectral profile obtained for UO$_2$Cl$_4^{2-}$. Our results underscore the importance of the equatorial ligands to simulating core spectra at both uranium and oxygen edges. 
\end{abstract}

\maketitle

\section{Introduction}
\label{sec:intro}
\noindent

Actinides are relevant to modern societies, first and foremost due to their central role in the nuclear industry~\cite{crossland2012nuclear,veliscek2016separation,mathur2001actinide,hu2022state,de2021extraction}, and the potentially harmful effects that can occur with their release in the biosphere in the case of industrial accidents~\cite{denecke2018sources,pavlyuk2022actinides,van2019quick,grambow2022ten}. In more recent years, the peculiar properties of materials containing actinides have also spurred interest from both a fundamental perspective as well as for industrial applications in catalysis~\cite{hu2018actinide,leduc2019chemistry,dennett2022towards}, materials~\cite{abney2017materials,parker2018overview,ji2020synchrotron,vallejo2022advances,leduc2019chemistry} and nanotechnology~\cite{zhai2022saxs,pallares2020transforming,ion2021green,bonato2020probing}.

As these properties stem from subtle differences in the electronic structure of different compounds~\cite{denning2007electronic,lv2019angle,sedigh2020review, wernet2019chemical}, a key objective in actinide science is to characterize their electronic structure under different conditions. To this end, spectroscopies involving core electrons~\cite{golubev2021core,stohr1992nexafs,van2016x,iwasawa2017xafs,stohr2019x} are particularly interesting since they are very sensitive to changes in oxidation state and on the local chemical environment of the actinides~\cite{vitova2018exploring,zegke2019differential,popa2016further,kvashnina2022high,vitova2010high}. 

In recent years, with the development of high-energy fourth-generation synchrotron radiation facilities around the world and the availability of X-ray Free Electron Lasers (XFELs) \cite{stohr2019x,van2016x,lindroth2019challenges}, one has witnessed the development and applications of High-Energy Resolution X-ray Spectroscopy (HERXS) to actinides~\cite{zimina2016cat,scheinost2021robl,caciuffo2021x,shi2018exploring,rothe2019fifteen,husar2022x}. 

The set of HERXS techniques includes spectroscopies such as High-Energy Resolution Fluorescence Detected X-ray Absorption Near Edge Structure (HERFD-XANES), High-Energy Resolution Auger Detection (HERAD-XANES), High-Energy X-ray Scattering (HEXS) and Extended X-ray Absorption Fine Structure (EXAFS)~\cite{vitova2013actinide,kvashnina2014role,kvashnina2022high,caciuffo2021x,hu2018actinide,husar2022x}.  For actinides, a particularly appealing feature of HERXS measurements is their larger penetration depth, which allows for smaller samples\cite{willmott2019introduction,iwasawa2017xafs,zimina2016cat,scheinost2021robl,caciuffo2021x,shi2018exploring}, this way minimizing the need for extensive radiation exposure protection measures. 
These experiments also offer increased sensitivity in probing processes that involve transitions to and from \textit{d} and \textit{f} orbitals, with the latter playing a pivotal role in actinide bonding ~\cite{nocton2021coordination,mcskimming2018coordination,vitova2015polarization,vitova2017role,gibson2018experimental,ganguly2020ab}.

These experimental developments, and the inherent complexities of interpreting the data generated, have stimulated the use of accurate molecular electronic structure methods as a means to interpret, predict and suggest new measurements in the X-ray range~\cite{norman2018simulating,besley2021modeling,zheng2019performance,wenzel2015analysis,vidal2019new,sen2018inclusion}.

While it has long been established that in this regime theoretical approaches must properly account for electron correlation\cite{south20164,zheng2019performance,park2020multireference,lischka2018multireference,wenzel2015analysis,norman1997geometry} and orbital relaxation\cite{gulania2021equation,vidal2020equation,pavoevi2020multicomponent,sen2018inclusion,krykunov2013self} in order to yield reliable results, the importance of relativistic effects  \cite{saue2011relativistic,liu2020essentials,dyall2007introduction,pyykko1978relativistic,reiher2014relativistic}, indispensable for heavy elements, is now also recognized for elements as light as those in the second row of the periodic table. 

Since most measurements of core spectra are done for molecules in the gas phase, it may become necessary to include environmental effects in the calculations. This can be done efficiently and accurately with embedding approaches~\cite{thiel2013methods,gomes2012quantum,jacob2014subsystem,jones2020embedding}, whereby a sufficiently accurate fully quantum mechanical (QM) description is applied to the region of interest and a significant portion of the system is represented via quantum (QM/QM) or classically-derived (QM/MM) effective operators, without any \textit{a priori} assumptions on the nature of the region of interest or the environment--which makes them potentially more flexible than other proven approaches based on e.g.\ explicit and continuum (implicit) solvation models \cite{di2014four,parmar2015applications,oher2020investigation,oher2021influence} and ligand field theory~\cite{karbowiak1997spectroscopic,thouvenot1994spectroscopic,butorin2022x}.

Among actinides, a particularly important class of complexes contains the uranyl (UO$_2^{n+}, n=1,2$) moiety, which is ubiquitous in the solution and solid-state chemistry of uranium. It is known to show very strong U-O triple bonds~\cite{denning2007electronic} and generally presents a linear O-U-O geometry, with other ligands coordinating with uranium in the equatorial plane, via less strong interactions. The core spectroscopy of uranyl coordination complexes has been investigated with several theoretical methods, ranging from complete and restricted active space self-consistent field~\cite{sergentu2018ab,ramanantoanina2019study,ganguly2020ab,polly2021relativistic,sergentu2022x,sergentu2022y}, crystal ligand-field multiplet theory and its variations \cite{tobin2022unoccupied,ilton2008ligand, ramanantoanina2016core, ramanantoanina2017calculation,vitova2010high,kvashnina2014role,butorin2022x}, density functional theory in its different flavors \cite{Spencer2013,pidchenko2017uranium,vitova2015polarization,kvashnina2014role,minasian2012determining,minasian2012determining,south20164,konecny2021accurate}, self-consistent real-space multiple scattering \cite{vitova2013actinide,kvashnina2014role}, static exchange approximation \cite{south20164} and perturbation theory \cite{south20164}. 

For molecular-based approaches, relativistic correlated many-body approaches~\cite{halbert2021relativistic,polly2021relativistic,sergentu2022x,sergentu2022y} are among the most accurate approaches that can be used, but their relatively high computational cost makes it difficult to employ for the experimentally relevant systems in the condensed phase. In the case of absorption spectroscopy, it has been found that with a suitable choice of density functional approximation (DFA) -- notably the CAM-B3LYP functional~\cite{yanai2004new} -- density functional theory (DFT) can provide reliable excitation energies of actinides in the ultraviolet-visible region ~\cite{tecmer2012charge,gomes2013towards,oher2020investigation,oher2021influence,real2007theoretical,oher2020influence}, with a recent, very comprehensive work by~~\citet{konecny2021accurate} employing the four component Damped Response Time-Dependent Density Functional Theory (4c-DR-TD-DFT) formalism, indicating that the same holds for core excited states of uranium complexes.

One aspect that, in our view, was not sufficiently addressed in the work by Konecny and co-workers, is the extent to which the environment can play a role in the spectral features. While~~\citet{konecny2021accurate} provided a comparison between theory and experiment for the uranium M$_4$-edge of uranium nitrate oxide  (UO$_2$(NO$_3$)$_2$), there are other systems such as the Cs$_{2}$UO$_{2}$Cl$_{4}$ crystal, for which there are HERFD-XANES spectra for the uranium M$_4$-edge as well as XANES spectra for the  uranium L$_3$-edge~\cite{vitova2015polarization}, and oxygen K-edge~\cite{denning2002covalency}, which could provide more comprehensive comparisons to experiment.

The present work, therefore, aims to employ embedding approaches and the 4c-DR-TD-DFT formalism to investigate  the effect of the chloride ligands bound to the equatorial plane of uranyl in the Cs$_{2}$UO$_{2}$Cl$_{4}$ crystal, on the oxygen K-edge and uranium M$_4$- and L$_3$-edges excited states, with a detailed comparison to experimental results available in the literature for this system. 

The structure of this article is as follows. We start by presenting the computational details from our calculations, followed by our results for the simulated X-ray absorption spectra. Finally, we present our conclusions and some future perspectives.

\section{Computational Details\label{sec:compdet}}

DR-TD-DFT and the corresponding 4c-DR-TD-DFT-in-DFT calculations were performed in DIRAC22~\cite{DIRAC22} version of the \texttt{DIRAC} electronic structure code~\cite{saue2020dirac}, as well as development snapshots (\texttt{34fbd49,4b35e48,d70bbe283,e061718,e0617189f6, \\ e7e2094}). We employed Dyall's all-electron basis sets of double and triple-zeta quality \cite{dyall2002relativistic,dyall2016relativistic} for uranium and Dunning's cc-pVTZ basis set \cite{denning1976electronic} for all other atoms. These basis sets were left uncontracted. We have employed a gaussian nuclear model in all calculations \cite{visscher1997dirac}.

For all DFT calculations with \texttt{DIRAC}, we have employed the 
Dirac-Coulomb ($^4$DC) Hamiltonian and the long-range corrected CAM-B3LYP\cite{yanai2004new} functional.

In all of our calculations, the structures employed were based on the experimental crystal structure of Cs$_2$UO$_2$Cl$_4$ reported by~~\citet{watkin1991structure}, whose U-O and U-Cl bond lengths are 1.774~\AA \ and 2.673~\AA \ respectively. For the embedding calculations, we employed the same structural models and subsystem partitioning as outlined by~~\citet{gomes2013towards}. All calculations on the bare uranyl and uranyl tetrachloride were carried out in D$_{\infty_h}$ and $D_{2h}$ symmetry, respectively, while for embedded uranyl simulations the crystalline site symmetry (C$_{2h}$) was used. 

The embedding potentials employed here were obtained from freeze-thaw calculations employing the scalar-relativistic ZORA Hamiltonian \cite{van1996zero}, TZ2P basis sets, the PW91k kinetic energy \cite{Lembarki1994}, the PBE exchange-correlation functional \cite{ernzerhof1999assessment} for the non-additive terms and subsystem energies. These calculations were carried out via the \texttt{PyADF} scripting framework \cite{jacob2011pyadf} and the embedding potentials were subsequently imported into the \texttt{DIRAC} calculations. 

For the 4c-DR-TD-DFT calculations we selected a frequency range that covered the energy ranges bracketing the main features in the oxygen K-edge (around 510-540 eV) and the uranium M$_4$- and L$_3$-edges (around 3680-3710 eV and 17070-17100 eV, respectively).  Given the expected shift in these simulations, the first frequency for each interval was obtained through TD-DFT simulations using the restricted-energy window scheme (REW-TD-DFT) \cite{zhang2012core} implementation on \texttt{DIRAC}.

In addition to the calculations with \texttt{DIRAC}, for the uranyl tetrachloride dianion species we have carried out standard (without embedding) TD-DFT calculations with the X2C Hamiltonian (including spin-orbit coupling, 2c-TD-DFT), triple-zeta basis functions with two polarization functions (TZ2P) \cite{van2003optimized}, and the CAM-B3LYP functional with the Amsterdam Density Functional (ADF) software package \cite{te2001chemistry}. As 4c-DR-TD-DFT calculations including spin-orbit coupling are not currently supported by ADF, we employed 2c-REW-TD-DFT calculations to target the core excited states of interest and employed the Tamm-Dancoff approximation (TDA) \cite{hirata1999time}, as it yielded spectra of the same quality as standard TD-DFT calculations at a lower computational cost. A comparison of TD-DFT and TDA can be found in Figures~\ref{fig:tda-vs-tddft} and~\ref{fig:4c2cntosuo2cl4} in the supplementary information.

Due to a lack of equivalent functionality in \texttt{DIRAC}, from the ADF 2c-REW-TD-DFT calculations we obtained natural transition orbitals (NTOs) \cite{martin2003natural}, which we used to provide a qualitative analysis of the changes to the electronic structure involved in the core excitations of uranyl tetrachloride. 

For simplicity, in the following, we shall refer to our 4c-DR-TD-CAM-B3LYP calculations as 4c-DR, and to our 2c-REW-TD-DFT and 2c-REW-TD-DFT-TDA CAM-B3LYP calculations as 2c-TD and 2c-TDA, respectively.


\section{Results and Discussion\label{sec:results}}

\begin{figure*}
    \centering
    \includegraphics[width=0.8\linewidth]{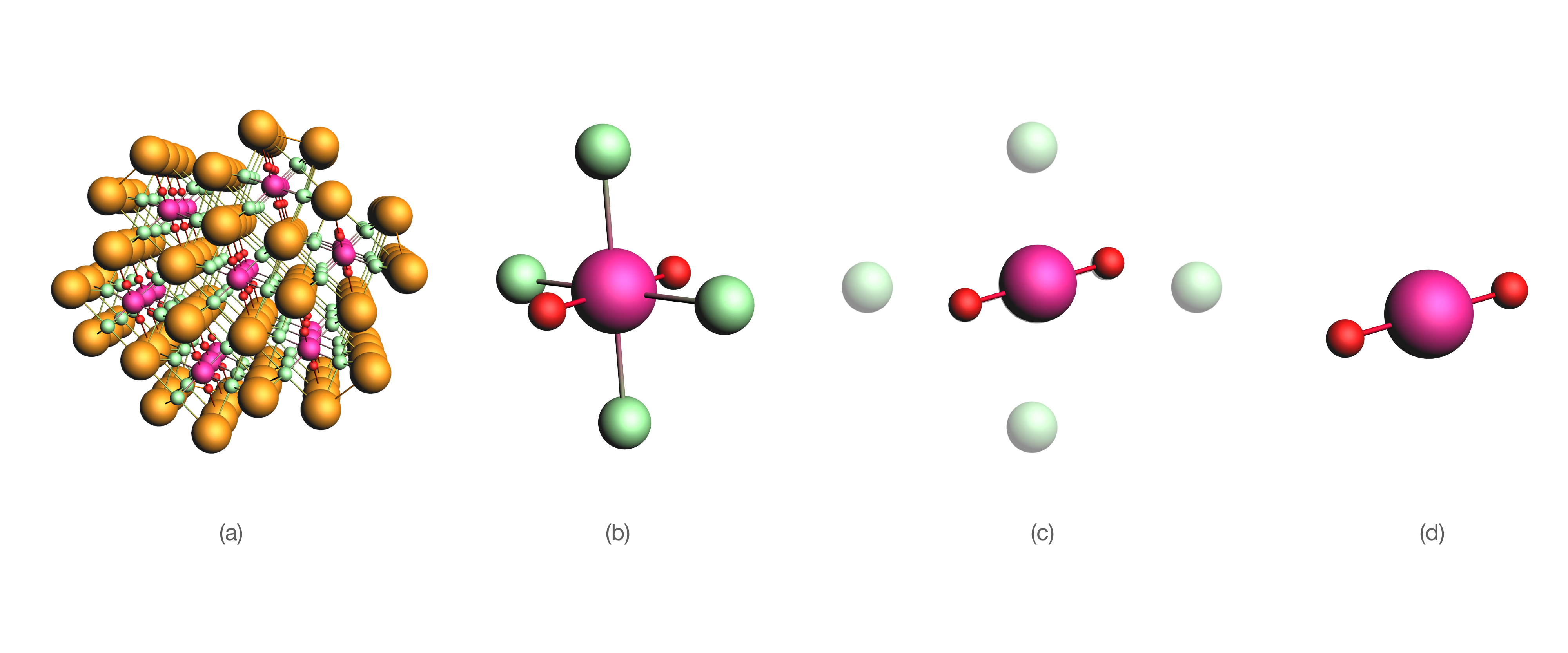}
    \caption{Reference system for this study: (a) Dicesium uranyl(VI) tetrachloride crystal (Cs$_2$UO$_2$Cl$_4$). Models here investigated: (b) uranyl(VI) tetrachloride dianion, UO$_2$Cl$_4^{2-}$ (c) uranyl(VI) ion in the FDE embedding potential of four chloride atoms, UO$_2^{2+}$ @ Cl$_4^{4-}$ and (d) bare uranyl(VI) ion, UO$_{2}^{2+}$ (cesium: orange; uranium: pink; oxygen: red; chlorine: green). See \citet{gomes2013towards} for further details on the structural models.}
    \label{fig:models}
\end{figure*}

In the following section, we present the theoretical absorption spectra at the oxygen K-edge and uranium M$_4$- and L$_3$-edges of the uranyl tetrachloride dianion (UO$_2$Cl$_4^{2-}$), a model in which the chloride ions bound to the equatorial plane of the uranyl ion (UO$_2^{2+}$) are represented by an FDE embedding potential (UO$_2^{2+}$ @ Cl$_4^{4-}$), and the uranyl ion without any chloride ligands (respectively structures \textbf{b}, \textbf{c} and \textbf{d} in Figure \ref{fig:models}). 

\begin{figure*}[!tbp]
  \centering
  \begin{minipage}[b]{\textwidth}
  \centering
  \includegraphics[width=\textwidth]{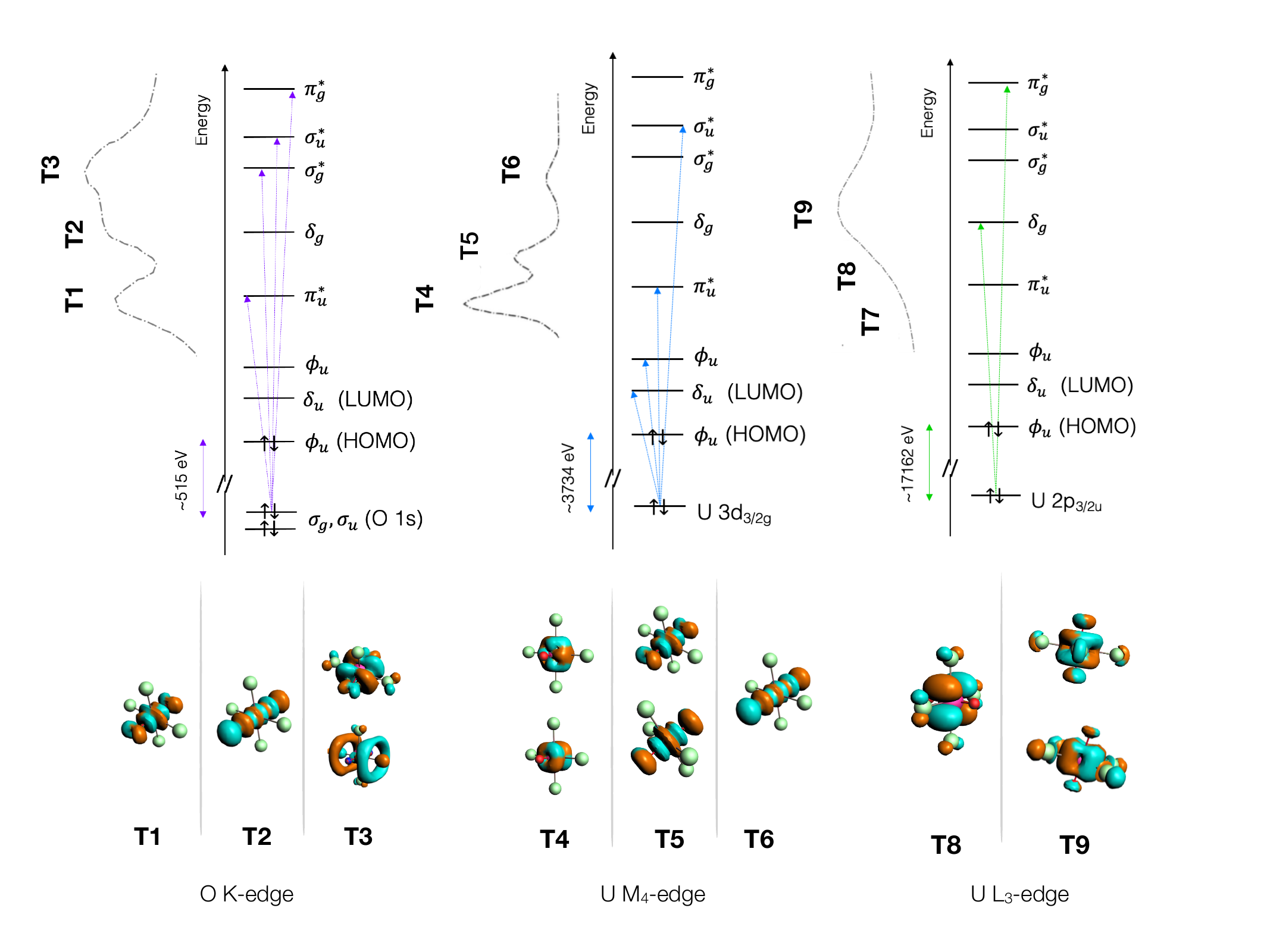}
  \end{minipage}
  \caption{Top: Qualitative MO diagram for the relevant orbitals of uranyl being accessed in this work. Bottom: Dominant 2c-TD natural transition orbitals (NTOs) for the peaks pertaining to the oxygen K-edge, and uranium M$_4$- and L$_3$-edges. Plots have employed 0.03 as the isosurface value. See the text for the nature of the transitions (labeled T1-T9). We do not display NTOs for T7 as the associated transition does not have intensity within the dipole approximation and is therefore not discussed.}
  \label{fig:ntos}
\end{figure*}

Before discussing the relative performance of these different models, we focus on a comparison between our results for UO$_2$Cl$_4^{2-}$ and the available experimental spectra recorded for the Cs$_2$UO$_2$Cl$_4$ crystal. Our results will be interpreted on the basis of the dominant NTOs for each peak, depicted in Figure~\ref{fig:ntos}, whereas the simulated spectra are depicted in Figure \ref{fig:poldep}.

\subsubsection{Features in the X-ray absorption spectra\label{sec:spec_lif}}

The absorption spectrum of the Cs$_2$UO$_2$Cl$_4$ crystal has been utilized in multiple studies as a tool to examine the electric properties of the uranyl(VI) ion in diverse environments, including one- and two-photon absorption at visible light \cite{denning1976electronic,barker1992applications}. The X-ray Absorption Near Edge Structure (XANES) spectrum at the oxygen K-edge of Cs$_2$UO$_2$Cl$_4$ was first reported by~~\citet{denning2002covalency} in the early 2000s, and the High Energy Resolution Fluorescence Detection (HERFD) spectra at the uranium M$_4$- and L$_3$-edges of Cs$_2$UO$_2$Cl$_4$ was reported by~~\citet{vitova2015polarization} in the past decade. 

As previously reported by~~\citet{denning2002covalency}, the O K-edge absorption spectrum (Figure \ref{fig:poldep} \textbf{a}) exhibits three prominent features  identified as (\textbf{T1}) a low-intensity pre-edge feature at 531.4 eV corresponding to an O 1s $\rightarrow$ $\pi_{u}^{*}$ transition, followed by (\textbf{T2}) an intermediate-intensity pre-edge feature around 534.1 eV, arising from an O 1s $\rightarrow$ $\sigma_{u}^{*}$ transition, and (\textbf{T3}) the white-line feature at 536.5 eV, where an O 1s $\rightarrow \pi_{g}^{*}$ transition is observed.

Using the HERFD mode for the evaluation of the U M$_4$-edge spectra it is possible to achieve experimental spectral widths below the natural core-hole lifetime broadening of approximately 4 eV. As a result, the high-resolution U M$_4$-edge spectra in Cs$_2$UO$_2$Cl$_4$ exhibit three well-defined features (Figure \ref{fig:poldep} \textbf{b}), which have been reported in previous studies \cite{vitova2015polarization,kvashnina2022high}: (\textbf{T4}) a U 3d$_{5/2g}$ $\rightarrow$ 5f${\delta_{u}}$ and U 3d$_{5/2g}$ $\rightarrow$ 5f${\phi_{u}}$ excitation observed at 3726.4 eV, (\textbf{T5}) a U 3d$_{5/2g}$ $\rightarrow$ 5f${\pi_{u}^{*}}$ resonance at 3728.6 eV, and a satellite peak (\textbf{T6}) U 3d$_{5/2g}$ $\rightarrow$ 5f${\sigma_{u}^{*}}$ at 3732.3 eV.

Interpreting the spectra at the U L$_3$-edge (Figure \ref{fig:poldep} \textbf{c}) remains challenging due to the significant core-hole lifetime width spanned in this absorption edge, which extends between 7.4 -- 8.4 eV \cite{campbell2001widths,krause1979naturcal}. This results in a significant portion of the spectral content being obscured. 

The first transition in this spectra is a 2p$_{3/2u}$ $\rightarrow$ 5f quadrupolar transition (\textbf{T7}) at 17168.8 eV, which is absent in our simulations. This is due to our use of the dipole approximation so that by symmetry the associated transition does not carry any intensity. Beyond that, the experimental white line is observed at 17175.2 eV, and our simulations predict that this feature it is mainly composed of (\textbf{T8, T9}) two 2p$_{3/2u}$ $\rightarrow$ (6d${\delta_g}$,6d${\pi_g}$) contributions, with a separation of 3.5 eV.

From the NTO analysis we see that the dominant contributions involve low-lying virtual orbitals centered around the uranyl moiety, and as discussed below the associated excited states will be rather well described. That said, one should be careful if highly diffuse orbitals such as Rydberg orbitals~\cite{Reisler2009,Frati2020} are found to be involved in the transitions, since apart from the general difficutly of describing such orbitals with standard gaussian basis sets~\cite{Kaufmann1989}, not all density functional approximations can describe long-range interactions as accurately as CAM-B3LYP~\cite{tecmer2012charge}.

\subsubsection{Theoretical vs experimental peak positions}
 
\begin{figure*}[!tbp]
  \centering
  \begin{minipage}[b]{\textwidth}
  \centering
  \includegraphics[width=0.90\textwidth]{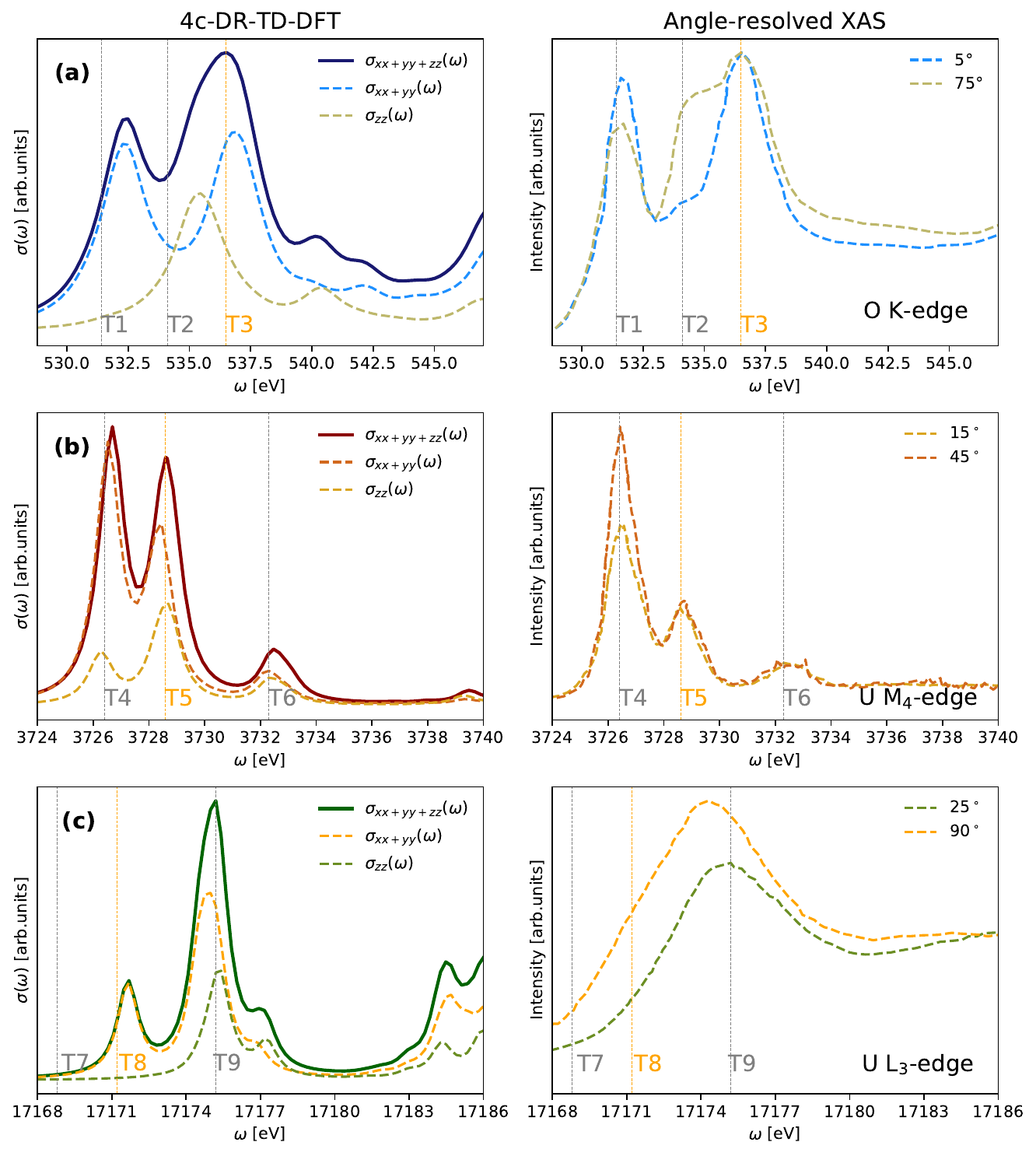}
  \end{minipage}
    \caption{Left: Partial (dashed) and total (solid) contributions to total oscillator strengths in the 4c-DR XAS spectra at (a) oxygen K-edge and uranium (b) M$_4$- and (c) U L$_3$-edges of UO$_2$Cl$_4$$^{2-}$. Right: Experimental data (Cs$_2$UO$_2$Cl$_4$) at the O K-edge (digitized with permission from~\citet{denning2002covalency}. Copyright 2002 AIP Publishing) and U M$_4$-, L$_3$-edges (digitized from~\citet{vitova2015polarization}. Copyright 2015 American Chemical Society). The angle indicated in the figures corresponds to the angle between the incident light beam and the O-U-O axis, respectively.}
    \label{fig:poldep}
\end{figure*}

All of our calculated transition energies shown in Table~\ref{tab:transition-energies} were considered for the analysis. We observe these are shifted in a very systematic manner to higher energies with respect to the experiment, with these differences of energy becoming increasingly pronounced at higher absorption edges. For instance, at the O K-edge, the energy shift values for the 2c-TDA and 4c-DR simulations are respectively 13.3 eV and 13.8 eV, whereas, for the U L$_3$-edge, these values are 95.2 eV and 86.3 eV. Therefore, in order to compare the theoretical and experimental results, we have aligned these based on specific transitions, identified by orange vertical lines in Figure~\ref{fig:poldep}. 

\begin{table}[h!]
\centering
\caption{Transition energies (in eV) calculated using 2c-TDA and 4c-DR compared to experimental values for the O K-edge, U M$_4$-edge, and U L$_3$-edge. Shifts with respect to experiment (in eV) are shown in parenthesis.}
\label{tab:transition-energies}
\begin{tabular}{ c c c c }
\hline
\multicolumn{4}{ c }{O K-edge} \\
\hline
 & T1  & T2 & T3 \\
 & O 1s $\rightarrow$ $\pi^{*}_{u}$ & O 1s $\rightarrow$ $\sigma^{*}_{u}$ & O 1s $\rightarrow$ $\pi^{*}_{g}$ \\
\hline
2c-TDA & 518.6 (-12.8) & 521.4 (-12.7) & 523.2 (-13.3) \\
4c-DR &  518.5 (-12.9) & 521.4 (-12.7) & 522.7 (-13.8) \\
Exp.~\cite{denning2002covalency} & 531.4 & 534.1 & 536.5 \\
\hline
\multicolumn{4}{ c }{U M$_4$-edge} \\
\hline
 & T4  & T5 &  T6 \\
 & U 3d$_{5/2g}$ $\rightarrow$ 5f${\delta_{u}}$ & U 3d$_{5/2g}$ $\rightarrow \pi^{*}_{u}$ & U 3d$_{5/2g}$ $\rightarrow$ 5f${\sigma^{*}_{u}}$ \\
\hline
2c-TDA & 3686.6 (-40.8) & 3688.4 (-40.2) & 3692.5 (-39.8) \\
4c-DR  & 3689.5 (-36.9) & 3691.4 (-37.2) & 3695.3 (-37.0) \\
Exp.~\cite{vitova2015polarization} & 3726.4 & 3728.6 & 3732.3 \\
\hline
\multicolumn{4}{ c }{U L$_3$-edge} \\
\hline
 & T7 & T8 & T9 \\
 & U 2p$_{3/2u}$ $\rightarrow$ 5f & U 2p$_{3/2u}$ $\rightarrow$ 6d${\delta_g}$ &  U 2p$_{3/2u}$ $\rightarrow$ 6d${\pi_g}$\\
\hline
2c-TDA & - & 17076.5 (-95.2) & 17080.0 (-95.2) \\
4c-DR  & - & 17085.4 (-86.3) & 17088.9 (-86.3) \\
Exp.~\cite{vitova2015polarization} & 17168.8 & 17171.7 & 17175.2 \\
\hline
\end{tabular}
\end{table}

The increasing discrepancy between experimental and theoretical peak positions (and hence the increase in the value of the shift to be applied to theoretical results) for deeper cores is related to the difficulty of TD-DFT to properly account for the orbital relaxation that should accompany the creation of a core hole--which will be increasingly important as the core orbitals under consideration become closer to the nucleus. 

The magnitude of orbital relaxation effects for the deeper core in heavy elements is clearly illustrated for ionization energies, obtained for small actinide species~\cite{south20164} with state-specific correlated methods ($\Delta$MP2) or for heavy halogenated species~\cite{southworth2019observing,halbert2021relativistic,Knecht2022} employing equation of motion coupled-cluster approaches (CVS-EOM-IP-CCSD), with the results by~\citet{halbert2021relativistic} and~\citet{Knecht2022} illustrating that (small) differences between 2- and 4-component Hamiltonians as seen here may arise, depending on which 2-component approach is used.  More recently, orbital relaxation has also been investigated for excitation energies at the U M$_4$-edge of the bare uranyl using multiconfigurational approaches ~\cite{sergentu2018ab,polly2021relativistic,sergentu2022x}.

In the latter case,~~\citet{polly2021relativistic} evaluated the performance of various active spaces in SO-RASSCF and SO-RASPT2 simulations. Their results to excitation energies at the U M$_4$-edge were found to overestimate experimental values by 12.9 eV and 17.4 eV at RASSCF and RASPT2 theory levels, respectively. This corresponds to about half of the overestimation found for our 4c-DR and 2c-TDA results (37.2 eV and 39.8 eV, respectively). 

The relatively modest changes upon including dynamical electron correlation (about 5 eV) are consistent with the findings of~~\citet{south20164} on the contribution from electron correlation (at MP2 level) to the binding energies in uranyl, as 
most of the core hole relaxation is coming from the orbital optimization enabled by the RASSCF approach. It is nevertheless interesting to note that the RASSCF calculations are closer to the experiment than RASPT2, suggesting a possible cancellation of errors when comparing the bare uranyl to the experimental system. We will return to this point when discussing the performance of embedding.

As final remarks, we note that the damping factors employed for determining the 4c-DR absorption spectra shown in Figure~\ref{fig:poldep} are not arbitrary but were selected to best reflect the profile observed in the experimental data for the Cs$_2$UO$_2$Cl$_4$ crystal, after exploring different values for each edge.  For the oxygen K-edge, the value is somewhat larger ($\gamma=1.0$ eV) than for the uranium M$_4$- and L$_3$-edges ($\gamma=0.5$ eV), but nevertheless our results show that with these choices the key features of both conventional and high-resolution spectra can be clearly identified. 

Furthermore, the broadening profile presented in Figure \ref{fig:poldep} was chosen to better fit the experimental data: for the O K-edge and  U L$_3$-edge spectra, we employed a Voigt profile, and Gaussian function for the U M$_4$-edge. In the supplementary information, Figure~\ref{damping-1} presents the effect of the damping factors, and Figures~\ref{broadening-1},~\ref{broadening-2}, and~\ref{broadening-3} the effect of the broadening profiles on the simulated spectra.

\subsubsection{Polarization dependence of calculated intensities}

Synchrotron-based experiments can also handle dichroism in the X-ray regime. Linear dichroism is studied using angle-resolved and polarization-dependent X-ray absorption spectroscopy~\cite{vitova2015polarization,denning2002covalency,frati2020oxygen}, 
while magnetic circular dichroism has been extensively studied through X-ray magnetic circular dichroism (XMCD) \cite{ogasawara1991calculation,van2014x}. 

By utilizing the damped-response theory, linear dichroism can be explored through the analysis of the components of the complex polarizability tensor ($\alpha_{xx}(\omega)$, $\alpha_{yy}(\omega)$, $\alpha_{zz}(\omega)$).
As we shall see in the following, the breakdown into individual components--or here, due to the symmetry of the system into parallel ($\sigma_{zz}$) and perpendicular ($\sigma_{xx+yy} = \sigma_{xx}+\sigma_{yy}$) components with respect to the uranyl bond axis--is helpful to understand the origin of the asymmetries observed experimentally.

In Figure \ref{fig:poldep} we show the parallel and perpendicular contributions to the total absorption cross-section ($\sigma_{xx+yy+zz} = \sigma_{zz} + \sigma_{xx+yy}$) for the edges under investigation, alongside the experimental polarization-dependent spectra~\cite{denning2002covalency,vitova2015polarization}, in which the angle indicated is the one between the incident light beam and the O-U-O axis in the Cs$_2$UO$_2$Cl$_4$ crystal.

Effects of polarization dependence are particularly prominent in the experimental spectra at the O K-edge by~\citet{denning2002covalency} and U L$_3$-edge by~\citet{vitova2015polarization} (Figure \ref{fig:poldep} \textbf{a} and \ref{fig:poldep} \textbf{c} respectively), which were recorded at greater incidence angles than the ones for the U M$_4$-edge. For clarity, here we will compare our results to the experiments under grazing and nearly perpendicular 
light incidence conditions. 

For the O K-edge, our analysis finds that transitions \textbf{T1} and \textbf{T3} are dominated by perpendicularly polarized components with respect to the O-U-O axis, whereas \textbf{T2} was found to exhibit parallel polarization, in agreement with experiment. For the L$_3$-edge, we observe that \textbf{T8} is completely dominated by perpendicularly polarized components. For \textbf{T9}, the spectrum shows non-negligible contributions from the parallel component, which we consider to be consistent with the observation by~\citet{vitova2015polarization} that angle dependence shifts the U L$_3$-edge absorption edge to the right when incoming light is parallel to the O-U-O axis.

Finally, for the U M$_4$-edge (Figure \ref{fig:poldep} \textbf{b}), a strong angular dependence in the white line (\textbf{T4}) is observed, and as for the other edges, the perpendicular components are the major contributors to the intensity. For \textbf{T5} we see that the parallel component contributes proportionally more to the total signal, and this trend is accentuated for \textbf{T6}, for which we see nearly equal contributions from both components to the total intensity. 

\subsubsection{A closer look on the role of the equatorial ligands\label{sec:dr_models}}

\begin{figure}[!tbp]
  \centering
  \begin{minipage}[b]{\columnwidth}
    \centering
    \includegraphics[width=0.8\textwidth]{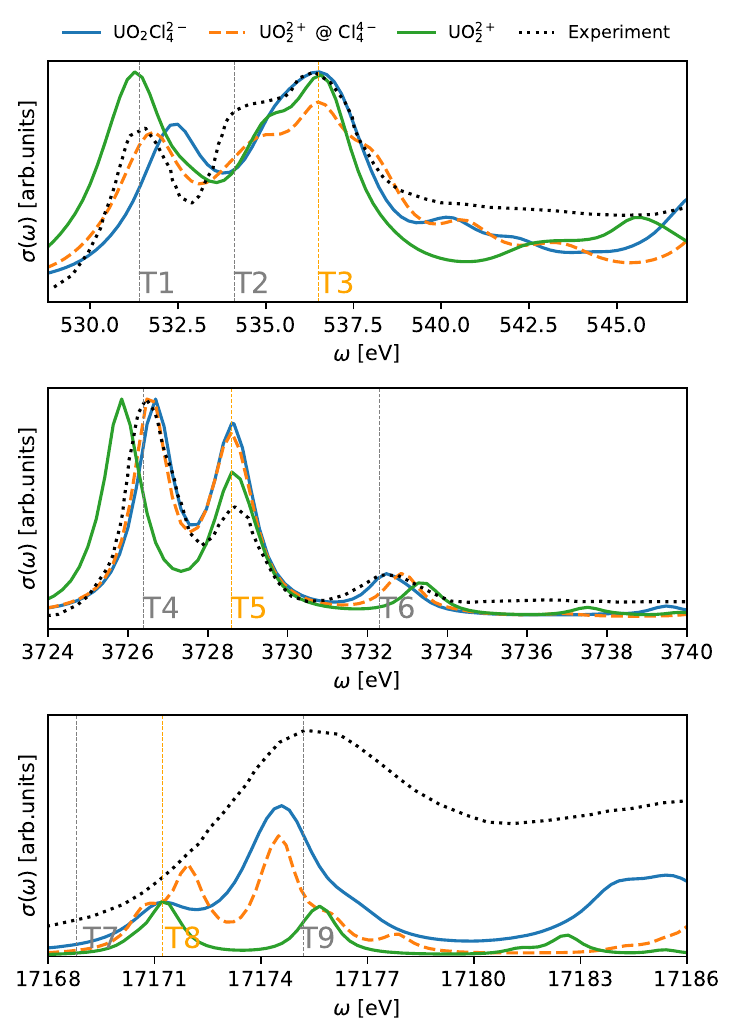}
  \end{minipage}
  \caption{From top to bottom: Comparison of the 4c-DR XAS spectra at the oxygen K-edge, uranium M$_4$ and L$_3$-edges of UO$_{2}^{2+}$, UO$_2^{2+}$ @ Cl$_4^{4-}$ and UO$_{2}$Cl$_{4}^{2-}$. Experimental data at the O K-edge in Cs$_2$UO$_2$Cl$_4$ (digitized with permission from~\citet{denning2002covalency}. Copyright 2002 AIP Publishing), and for the U M$_4$- and L$_3$-edges (digitized from~\citet{vitova2015polarization}. Copyright 2015 American Chemical Society), respectively.} 
  \label{fig:dr-td_models}
\end{figure}

Having established that 4c-DR and 2c-TDA can faithfully describe the spectra for the edges under consideration, we can turn to a comparison of the different structural models described in Figure~\ref{fig:models} (UO$_2$Cl$_4^{2-}$, the bare uranyl ion and the uranyl ion embedded onto the equatorial chloride ligands) in order to gain further understanding on the role of the equatorial ligands. 

As mentioned above, the core spectra of bare uranyl ion have been investigated by~\citet{polly2021relativistic}, but also by~\citet{sergentu2018ab}. The embedded uranyl model (UO$_2^{2+}$ @ Cl$_4^{4-}$), on the other hand, has only been previously investigated for valence excited state~\cite{gomes2013towards} and shown to yield a rather good description of the low-lying states of UO$_2$Cl$_4^{2-}$ in Cs$_2$UO$_2$Cl$_4$. 

Given the high sensitivity of core excited states to the description of the environment around the absorbing sites, it remains to be seen to which extent the embedded model can faithfully reproduce the reference (UO$_2$Cl$_4^{2-}$) calculations. In this respect, an interesting feature to track is the relative positions of the different peak maxima as well as their relative intensities, and how these compare to the experimental ones.
 
At the O K-edge (Figure \ref{fig:dr-td_models}, top), there are rather small differences between the energy shifts (less than 1 eV) with respect to the experiment among the three simulations. This is in line with a systematic error in the calculations due to the shortcomings in describing relaxation due to the creation of the core hole, mentioned in the previous section. 

For the intensities, on the other hand, we observe for \textbf{T1} that the bare uranyl model shows a significant overestimation. Intensities for the reference and embedded models on the other hand are very similar to one another, though the maximum for the embedded model is slightly shifted to lower energies. For \textbf{T2} and \textbf{T3}, the different models show an overall good agreement to each other, and a finer comparison is more difficult than for \textbf{T1} in view of the overlapping contributions from different transitions to the final spectrum.  We note however that for the bare and embedded uranyl the feature corresponding to the parallel component (\textbf{T2}, see Figure~\ref{fig:poldep} \textbf{a}) appears slightly more shifted to the lower energies than for uranyl tetrachloride, resulting in the asymmetry in the region between \textbf{T2} and \textbf{T3}. We note that experimentally the intensity of absorption between \textbf{T2} and \textbf{T3} is rather sensitive to the polarization\cite{denning2002covalency} (see Figure~\ref{fig:poldep} \textbf{a}). 

The simulations at the U M$_4$-edge spectra (Figure \ref{fig:dr-td_models}, middle) show an even better agreement between the embedding calculations and the reference ones than for the O K-edge, with both simulations reproducing the experiment rather well for all peaks. However, if the splitting between \textbf{T4} and \textbf{T5} is very well reproduced, we see that for the embedding calculation \textbf{T6} there is a slight shift in peak maximum to higher energies compared both to experiment and to the reference model while the latter two show again very good agreement. For the bare uranyl, on the other hand, we see a fairly significant overestimation of both \textbf{T4}-\textbf{T5} and \textbf{T5}-\textbf{T6} peak shifts compared to the reference model. 

The  ability of a theoretical model to accurately characterize peak splittings is of importance for the U M$_4$-edge since the solid-state community commonly attributes variations in peak splitting between the main feature and the satellite 5f$_{\sigma^{*}}$ in actinyl spectra to changes in overlap-driven covalency from variations in the An-O bond length, as suggested in the works of~~\citet{kvashnina2022high}, and Vitova and collaborators \cite{vitova2015polarization,vitova2017role}. This is somewhat at odds with the analysis based on \textit{ab initio} simulations, such as those by~~\citet{sergentu2018ab} and also~~\citet{polly2021relativistic}, which claim that there is not a clear correlation between those and the observed splitting in this spectrum.  The natural localized molecular orbitals (NLMOs) analysis by~~\citet{sergentu2018ab} found a significant decrease in the $\sigma$ covalency in the excited states of the actinyl absorption spectra at the M$_4$-edge of PuO$_2^{2+}$, NpO$_2^{2+}$, and UO$_2^{2+}$, in a trend that opposes the one observed in their ground states. Therefore, they were able to highlight the limitations of the previous interpretations of these spectra, by showing that the orbitals of the ground state should not be used as the sole guide in interpreting these actinyl spectra. 

\begin{table*}[]
\caption{U M$_4$-edge peak positions and differences (in eV) from theoretical molecular (4c-DR, RASPT2 and RASSCF) and finite-difference (FDMNES) calculations, and experiment (HERFD).}
\centering
\begin{tabular}{ccccccccc}
\hline\hline
& 
& \multicolumn{3}{c}{peak positions}
&
& \multicolumn{3}{c}{differences} \\
\cline{3-5}
\cline{7-9}
Method     
& System    
& \textbf{T4} 
& \textbf{T5}
& \textbf{T6}
&& \textbf{T5-T4}
& \textbf{T6-T4} 
& \textbf{T6-T5} \\ 
\hline
4c-DR & UO$_2^{2+}$ &  3689.0   & 3691.8  & 3696.5   && 2.8 & 7.5   & 4.7  \\
RASSCF~\cite{sergentu2018ab} & UO$_2^{2+}$ &  3750.1  & 3752.4  & 3758.0 && 2.3 & 7.9 & 5.6 \\ 
RASSCF$^a$~\cite{polly2021relativistic} & UO$_2^{2+}$ &  3748.0  & 3750.1  & 3755.7   && 2.1 & 7.7 & 5.6 \\ 
RASSCF$^b$~\cite{polly2021relativistic} & UO$_2^{2+}$ &  3750.6 & 3753.1  & 3759.4   && 2.5 & 8.8 & 6.3 \\ 
RASPT2$^b$~\cite{polly2021relativistic} & UO$_2^{2+}$ &   3746.1
&  3748.8 &  3755.1 && 2.7 & 9.0 & 6.3 \\
&&&&&&&&\\
4c-DR & UO$_2^{2+}$ @ Cl$_4^{4-}$  & 3690.1  & 3692.0 & 3696.2  && 1.9 & 6.1  & 4.2 \\
4c-DR & UO$_2$Cl$_4^{2-}$ & 3689.5  & 3691.4  & 3695.3  && 1.9 & 5.8  & 3.9 \\ 
&&&&&&&&\\
FDMNES~\cite{Amidani2021} & Cs$_2$UO$_2$Cl$_4$ & 3726.9 & 3729.3 & 3732.9 && 2.4 & 6.0 & 3.6 \\
HERFD~\cite{vitova2015polarization}  & Cs$_2$UO$_2$Cl$_4$ & 3726.4  & 3728.6 & 3732.3 && 2.2 & 5.9  & 3.7  \\
\hline\hline
\multicolumn{9}{l}{Active space: $^{a}$ 3d ($\sigma_{u}$,$\pi_{u})$/5f, $^{b}$ 3d/5f }\\
\end{tabular}
\label{table*:um4peaks}
\end{table*}

In Table ~\ref{table*:um4peaks}, we present the peak positions and splitting of the U M$_4$-edge spectra obtained from experiments by~~\citet{vitova2015polarization} and predicted by molecular \textit{ab-initio} calculations (\citet{sergentu2018ab},~~\citet{polly2021relativistic} and this work) as well as \textit{ab-initio} finite-difference calculations by~\citet{Amidani2021} using the FDMNES code~\cite{Bunu2021,Joly2022}. 

As previously discussed, the shortcomings of 4c-DR calculations in including relaxation result in excitation energies that are too low compared to experiment or RASSCF/RASPT2.  The peak splitting values, on the other hand, indicate that despite better accounting for orbital relaxation, RASSCF and RASPT2 calculations do not compare more favorably to the experiment than 4c-DR calculations relying on the TDDFT/CAM-B3LYP, and changing the size of the active space changes peak positions by a few eV but peak splittings by a few tenths of eV. Relaxation aside, the comparable performance of CAM-B3LYP with respect to RASSCF/RASPT2 is in line with previous benchmark studies~\cite{Tecmer2011,Tecmer2014}, and has been explained in terms of its good ability to minimize self-interaction errors that plague other functionals, so that excitation energies can be reliably obtained with it, including for charge-transfer excitations\cite{tecmer2012charge,Oher2023}.

Taken as a whole we attribute the overestimation of peak splittings shown for all of these calculations to the lack of equatorial ligands since when these are included (embedding or supermolecular 4c-DR calculations) we achieve  very good agreement with the experiment, especially for the  \textbf{T6-T4} and \textbf{T6-T5} splittings. It is interesting to note that the RASPT2 approach, which one could expect to improve upon RASSCF, ends up further overestimating peak splittings. This could be due to a poor balance between orbital relaxation and the amount of dynamic correlation recovered with second-order perturbation theory, possibly coupled with the lack of equatorial ligands in the calculations.  

Coming back to the  \textbf{T6-T4} and \textbf{T6-T5} splittings, we observe slight differences between the embedding and supermolecule calculations, which would suggest that orbital interactions are somewhat more important for \textbf{T6} than for \textbf{T4} or \textbf{T5}, since for the  \textbf{T5-T4} splitting for both models is essentially the same, whereas it increases slightly for \textbf{T6-T5}.

In the embedding calculation, apart from the symmetry breaking with respect to linear symmetry than allows for some mixing of the $\phi, \delta$ spinors~\cite{Gomes2008}, the chloride ligands are included only through the embedding potential, which represents an effective interaction. As a result, by construction, the electrostatic interaction component of the embedding potential remains the dominant contribution. Therefore, the differences between the two models can be traced back to the lack of such interaction in the bare uranyl case. 

The importance of electrostatic interactions, and the sensitivity of the actinyl electronic structure to these, has been addressed in more detail in our previous investigations of actinyl chlorides. First, for the valence states of uranyl~\cite{gomes2013towards}, we have also considered a simple model with negative point charges placed in the positions of the chlorides, and found that the valence excited states were in better agreement with uranyl tetrachloride than the bare uranyl, though this point-charge embedding model still showed a poorer performance than the FDE (UO$_2^{2+}$ @ Cl$_4^{4-}$) model. Second, in the  investigation~\cite{Gomes2008} of $f$-$f$ transitions in NpO$_2$Cl$_4^{2-}$, we considered an embedding model equivalent to the one used here (NpO$_2^{2+}$ @ Cl$_4^{4-}$), as well as a variant in which the density of the chlorides was not relaxed in the presence of the neptunyl ion, and a point-charge embedding model. We have shown the embedded model without relaxation of the chloride ligands' density yielded results which were comparable to the point charge one and did not reproduce the spectra of neptunyl chloride as well as NpO$_2^{2+}$ @ Cl$_4^{4-}$.

From the above considerations and the our current results, we argue that substantial contributions to the shifts in U M$_4$-edge peak position in table~\ref{table*:um4peaks} can be attributed to electrostatic interactions with the equatorial ligands.  
It would be interesting to investigate to which extent that would hold for the core spectra of other actinyls, and that for ligands other than halides, but such a study is beyond the scope of this work. In the case of other actinyls, which may present a multi-reference ground-state, it will be in general necessary to replace the DFT treatment of the active subsystem by a wavefunction-based method, but as we have shown for neptunyl~\cite{Gomes2008}, that does not pose particular issues for the FDE calculations. 

At the U L$_3$-edge (Figure \ref{fig:dr-td_models}) we have obtained the same results as~\citet{south20164} for the uranyl ion, which presents a second feature (\textbf{T9}) quite shifted to the right, therefore not being compatible with any spectra recorded at this absorption edge for uranium-containing species. The embedding calculation presents a broader and more intense \textbf{T8} feature, with the \textbf{T9} resembling rather well that of the reference model.

We note that here we have not considered more sophisticated models in which the crystal environment in Cs$_2$UO$_2$Cl$_4$ beyond the equatorial chloride ligands (structure \textbf{a} in Figure \ref{fig:models}) are included in the embedding potential.  This is due the fact that in our previous investigation of valence excited states~\cite{gomes2013towards} we have compared the results for the embedded uranyl tetrachloride model employed here (UO$_2^{2+}$ @ Cl$_4^{4-}$) with others in which the effect of the rest of the crystal was taken into account by embedding potentials. In these potentials we defined a frozen density region, consisting of the equatorial chlorides plus 20 uranyl tetrachloride and 90 cesium atoms surrounding the central uranyl unit; and beyond this quantum mechanical region, an array of point charges was used to simulate the long-range electrostatics in the crystal. 

For a first model, in which all subsystem densities but that of the chloride equatorial ligands were kept frozen, we observed for the 12 lowest valence excited states changes in excitation energies that differed at most by 0.015 eV (with discrepancies being on average of about 0.0075 eV, in absolute value) from the UO$_2^{2+}$ @ Cl$_4^{4-}$ model. Furthermore, when the density of of the nearest cesium atoms was also relaxed, the excitation energies only changed by around 0.002-0.004 eV. 

As discussed by~\citet{gomes2013towards}, the physical process behind these changes is that the effect of the crystal environment lowers the energies of both occupied and virtual orbitals by roughly the same amounts. In addition to that, the energies of occupied valence~\cite{Gomes2008,gomes2013towards,Bouchafra2018} and core levels~\cite{Opoku2022} are shifted by essentially the same amount by the environment. With that, relative measures such as excitation energies turn out to be weakly affected by the environment. 

Apart from the arguments above, we consider the suitability of our uranyl tetrachloride models to be further supported by the good agreement between all of our peak splittings for the uranyl tetrachloride models and those from~\citet{Amidani2021}, obtained with the FDMNES method. 

In the latter calculation, all atoms within a 6~\AA~radius around the absorbing uranium atom in the Cs$_2$UO$_2$Cl$_4$ crystal are take into account, which is in between our models without and with accounting for the rest of the crystal. We note that the FDMNES calculations are also based on DFT, and account for scalar and spin-orbit relativistic effects, but differ qualitatively from ours in that they include the effect of core hole creation through a screening parameter. With that, their absolute peak positions are also in good agrement with experiment, while ours require a shift in energy to be compared to experiment. 

For ionizations on the other hand long-range interactions with the environment will be important~\cite{gomes2013towards,Bouchafra2018,Opoku2022}, and will require the use of more sophisticated structural models. We shall address the ionizations of uranyl in Cs$_2$UO$_2$Cl$_4$ in a separate publication.

\section{Conclusions\label{sec:conclusions}}

\noindent

We have carried out an evaluation of relativistic quantum chemistry approaches for the calculation of excitation energies at the oxygen K-edge and uranium M$_4$-, L$_3$-edges, for the UO$_2$Cl$_4^{2-}$ system which makes up the Cs$_{2}$UO$_{2}$Cl$_{4}$ crystal. Besides calculating on this system, we have also investigated the performance of two other structural models, the bare uranyl ion and the uranyl ion embedded (via the frozen density embedding method, FDE) in an environment made up by the chloride equatorial ligands.

For these excitation energies, the use of the Coulomb-attenuating functional CAM-B3LYP in four-component damped-response simulations produced results that are consistent with the features observed in both conventional XANES and HERFD experiments. Additionally, through the analysis of natural transition orbitals (NTOs) obtained from equivalent two-component time-dependent DFT calculations, we showed that the O K-edge spectra primarily provide insight into the low-lying antibonding states of $\pi_{u}^{*}$, $\sigma_{u}^{*}$ and $\pi_{g}^{*}$ symmetries centered in the uranyl unit. In contrast, the lowest energy orbitals accessed at the U M$_4$- and U L$_3$- absorption edges are the uranium 5f$\phi_{u}/\delta_{u}$ and 6d${\delta_{g}/\pi_{g}}$ non-bonding orbitals. These observations are in line with previous investigations.

We have determined that the bare uranyl model shows significant changes in spectral profiles in comparison to structural models containing equatorial ligands, notably in terms of relative intensities but also to some extent with respect to relative peak positions. This is seen for the uranium edges, to which the equatorial ligands are attached, but also for the oxygen K-edge. 

In the soft and tender X-ray regime (O K- and U M$_4$-edge, respectively), we have observed that our simulations for the embedded model yield both excitation energies and intensities in very good agreement with those obtained for the UO$_{2}$Cl$_{4}^{2-}$ ion, though in the hard X-ray regime (U L$_3$-edge) this approximate model still performs well but deviates more significantly from the results obtained for the anion. 

In conclusion, our study has provided valuable insights into the impact of equatorial ligands on the X-ray absorption spectra of the uranyl ion. Our findings emphasize the limitations of relying solely on a structural model consisting of either the bare uranyl ion or the ground state structure of the uranyl site, as these models do not accurately account for the influence of the equatorial ligands on peak positions. 

Furthermore, our work demonstrates the potential of using accurate embedded models, such as those that can be constructed with frozen density embedding, to semi-quantitatively simulate actinyl core spectra.  FDE-based embedding models have already been employed with multireference wavefunction-based treatment for the active subsystem, to treat systems which are challenging for DFT or other single-reference methods, though not yet for core spectra of actinides employing accurate ab initio relativistisc quantum chemistry methods.

\begin{acknowledgement}

We acknowledge discussions with Dr.\ Valérie Vallet (Université de Lille) on carrying out the analysis of natural transition orbitals and on the manuscript. WAM acknowledges helpful discussions with Dr. Patrick Norman (KTH) on general aspects of response theory.
We acknowledge support from the Franco-German project CompRIXS (Agence nationale de la recherche ANR-19-CE29-0019, Deutsche Forschungsgemeinschaft JA 2329/6-1), PIA ANR project CaPPA (ANR-11-LABX-0005-01), I-SITE ULNE projects OVERSEE and MESONM International Associated Laboratory (LAI) (ANR-16-IDEX-0004), the French Ministry of Higher Education and Research, region Hauts de France council and European Regional Development Fund (ERDF) project CPER CLIMIBIO, and the French national supercomputing facilities (grants DARI A0090801859, A0110801859, A0130801859). 
\end{acknowledgement}

\section*{Associated content}
The data corresponding to the calculations of this paper are available at the Zenodo repository under DOI: \href{10.5281/zenodo.7632750}{https://doi.org/10.5281/zenodo.7632750} .

The Supporting Information is available free of charge on the \href{http://pubs.acs.org}{ACS Publications website} at DOI: \href{}{XXX}. It contains: (a) further theoretical background on damped response theory and frozen density embedding; (b) comparison of 2- and 4-component calculations for uranyl tetrachloride; (c) assessment of the effect of the damping paramenter used in thr damped response theory calculations; (d) a comparison of the effect of different broadening functions on spectral shapes; (e) a comparison on the differences in energy shifts with respect to experiment for the bare and embedded uranyl models, and uranyl tetrachloride. (PDF).

\bibliography{refs.bib}
\end{document}


\maketitle

\section{Theoretical Methods}
\label{sec:theory}

In this section, we recall the main features of the theoretical approaches used in this work.

\subsection{Damped-Response Theory}

The presence of a time-dependent external electric field with particular field strength $F(t)$ may induce electronic transitions in a molecular system. 

Within the electric-dipole approximation, the coupling between the electric dipole of the system ($\vec\mu$) and this external electric field produces the features observed in a spectrum. This way, perturbations on a system can be theoretically tracked by following the contribution of non-zero order terms on the field expansion of the time-dependent polarization:
\begin{equation}
 \mu(t) = \mu^{0} + \alpha F(t) + \frac{1}{2}\beta F^2(t) + \ldots
\end{equation} where $\mu^{0}$ denotes the permanent electric-dipole moment, $\alpha$ is the polarizability tensor, and $\beta$ is the first-order contribution to non-linear processes such as two-photon absorption (TPA). 

Within the framework of standard response theory, one-photon absorption processes can be studied by the computation of the resonant-divergent linear response function. In the exact theory case, the response function relating to the electric-dipole polarizability tensor can be expressed as a sum over states,
\begin{equation}
\small
\alpha_{ij}(\omega) = - \frac{1}{\hbar} \sum _{n >  0} \left [  \frac{\left \langle 0|\hat{\mu}_{i}|n \right \rangle \left \langle n|\hat{\mu}_{j}|0 \right \rangle}{\omega_{0n}- \omega}+ \frac{\left \langle 0|\hat{\mu}_{i}|n \right \rangle \left \langle n|\hat{\mu}_{j}|0 \right \rangle}{\omega_{0n}+ \omega}  \right ]
\label{def.standard_alpha} 
\end{equation} where $\hat{\mu}_{i}$ corresponds to a component of the electric-dipole operator along a particular direction, and $\omega_{0n}$ denotes the frequency associated with a transition between the ground state ($ | 0 \rangle $) and an excited state ($ | n \rangle $), and $\omega$ the frequency of the perturbing field.

In the approximate case, one generally refrains from using a sum over states expression but rather recasts the determination of the response functions in terms of the iterative solution of response equations and the subsequent assembly of the response functions.
From the linear response function, one obtains the excitation energies from the poles of the response function (i.e.\ when the perturbing frequency matches the one corresponding) while transition moments are obtained from the residues of the response function \cite{norman2018simulating,norman2018principles,norman2011perspective,kauczor2014efficient,christiansen1998response,helgaker2012recent}.

What Eq.~\ref{def.standard_alpha} misses, however, is accounting for the relaxation processes that yield finite lifetimes for the different excited states. Such relaxation can be incorporated into theoretical models in a phenomenological way by including damping terms $\gamma_{0n} = 1/\tau_{n}$ in the desired response function, where $\tau_{n}$ denote the finite lifetime a particular excited state ~\cite{norman2018simulating,norman2011perspective}. In this case, we have a redefinition in the denominator in Eq.~\ref{def.standard_alpha} such that 
\begin{equation}
\centering
\omega_{0n} \rightarrow \omega_{0n} \pm i\gamma_{0n}
\end{equation}
resulting in a polarizability tensor that becomes a complex quantity. This case is thus referred to as damped-response (DR) theory, for which 4- or 2-component based implementations have been presented for mean-field approaches such as DFT by~~\citet{villaume2010linear}--which is the implementation used in this work for all excitation energy calculations--and more recently by~~\citet{konecny2019relativistic,konecny2021accurate}.

Considering a randomly oriented molecular sample, the damped response absorption cross section is now written in terms of the isotropic average of the complex component of the polarizability
\begin{equation}
    \sigma(\omega)=\frac{\omega}{\varepsilon_0c}\Im m  \left [\frac{1}{3} (\alpha_{xx}(\omega) + \alpha_{yy}(\omega) + \alpha_{zz}(\omega))   \right ]
    \label{def.dr_alpha} 
\end{equation} where $\varepsilon_{0}$ denotes the vacuum permittivity and $c$ the speed of light.

As damped-response theory provides a resonant-convergent expression for the electric-dipole polarizability, one may perform a series of simulations across a range of selected polarizabilities around a resonance, and then extract a theoretical spectrum that contains information on the relaxation process, avoiding the need to introduce artificial broadening.

As previously suggested by~~\citet{fransson2013carbon}, the inclusion of a channel-restriction scheme in simulations of light-matter interaction in the X-ray range may circumvent the presence of undesired features due to the incompleteness of the basis set and the consequent discretization of the continuum. We also point out that any channel restriction can ease this type of simulation at high energy ranges, as the diagonalization procedure may become computationally infeasible when one includes all possible excitations on these. 

Due to that, the calculations carried out in our work employed the named restricted-excitation-window (REW), where projection operators ensure that excitations are restricted to those in which only the selected orbitals are involved, avoiding features from outer-electrons \cite{stener2003time,zhang2012core,besley2020density,besley2021modeling}.
 
\subsection{Frozen Density Embedding}

In the frozen density embedding (FDE) approach~\cite{wesolowski1993frozen,wesolowski2015frozen,gomes2012quantum,jacob2014subsystem} one recast DFT into a subsystem theory so that a given system can be partitioned into (at least) two distinct fragments-the active system, or subsystem $I$ (here a uranyl unit) and the environment or subsystem $II$ (the rest of the system, e.g. equatorial ligands). The total system's electron density is therefore rewritten as 
\begin{equation}
    \rho_{total}(\textbf{r})=\rho_{I}(\textbf{r})+\rho_{II}(\textbf{r})
\end{equation} 
while the energy
\begin{equation}
    E_{tot}[\rho_{I},\rho_{II}]= E_I[\rho_I] + E_{II}[\rho_{II}]+E_{int}[\rho_I,\rho_{II}]
\label{def.FDE_total_energy}
\end{equation}
is decomposed into subsystem energies $(E_i[\rho_i]; i = \text{I, II})$ and and interaction term $E_{int}[\rho_I,\rho_{II}]$,
\begin{equation}
\small
\begin{multlined}
    E_{int}[\rho_I,\rho_{II}]=\int \rho_I(\mathbf{r})v_{nuc}^{II}(\mathbf{r}) \ d^{3}r + \int \rho_{II}(\mathbf{r})v_{nuc}^{I}(\mathbf{r}) \ d^{3}r \\ + E_{nuc}^{I,II}  +\iint\frac{\rho_{I}(\mathbf{r})\rho_{II}(\mathbf{r'})}{|\mathbf{r}-\mathbf{r'}|}d^{3}r \ d^{3}r' \\ + E_{xc}^{nadd}[\rho_I,\rho_{II}] + T_{s}^{nadd}[\rho_I,\rho_{II}]
\end{multlined}
\end{equation} where $v_{nuc}^{i}$ denotes the nuclear potential associated with the atoms in $i$-th subsystem,  $E_{xc}^{nadd}$ and $T_s^{nadd}$ are respectively the non-additive exchange correlation and kinetic energies, defined as
\begin{eqnarray}
E_{xc}^{nadd}[\rho_I,\rho_{II}] &= E_{xc}[\rho_{total}] - E_{xc}[\rho_I] - E_{xc}[\rho_{II}] \\
T_{s}^{nadd}[\rho_I,\rho_{II}]  &= \tilde{T}_{s}[\rho_{total}] - \tilde{T}_{s}[\rho_I] - \tilde{T}_{s}[\rho_{II}] 
\end{eqnarray} 
with $ \tilde{T}_{s}$ representing an approximation to the non-interacting kinetic energy, calculated with density functional approximations such as that by~\citet{Lembarki1994}.

If all subsystems are treated at the Kohn-Sham DFT level, the individual subsystem energies $E_i[\rho_i]$ will be the same as those for the DFT formalism, and from minimization of the total energy (Eq. \ref{def.FDE_total_energy}) with respect to $\rho_I$ for a given (frozen) $\rho_{II}$ one can obtain the orbitals for subsystem $I$ by solving a set of Kohn-Sham-like equations 
\begin{equation}
\left [ -\frac{\nabla^{2}}{2}+v_{eff}^{KS}[\rho_I](\textbf{r})+v_{emb}^{I}[\rho_I,\rho_{II}](\textbf{r}) \right ]\phi_{k}^{I}(\textbf{r})=\varepsilon_{k}^{I}\phi_{k}^{I}(\textbf{r})     
\label{ksced}
\end{equation}
so that in the ground state the different subsystems are coupled to each other through the so-called embedding potential $v_{emb}^{I}$,
\begin{equation}
\begin{multlined}
    v_{emb}^{I}[\rho_I,\rho_{II}](\textbf{r})=\frac{\delta E_{int}[\rho_I,\rho_{II}]}{\delta\rho_I(\textbf{r})}=
v_{nuc}^{II}(\textbf{r})+ \\ \int \frac{\rho_{II}(\textbf{r}')d^{3}r '}{|\textbf{r}-\textbf{r}'|}+
\frac{\delta E_{xc}^{nadd}[\rho_I,\rho_{II}]}{\delta\rho_I(\textbf{r})}+
\frac{\delta T_{s}^{nadd}[\rho_I,\rho_{II}]}{\delta\rho_I(\textbf{r})}
\label{embpot}
\end{multlined}
\end{equation}
One can also optimize $\rho_{II}$, for instance via the freeze-thaw procedure in which the roles of $\rho_{I}$ and $\rho_{II}$ are exchanged when solving Eq.~\ref{ksced}. We found this to be particularly important for charged subsystems~\cite{Gomes2008,de2022frozen,Bouchafra2018}. 

In the case of a coupled cluster treatment of subsystem $I$, we can recast the equations above into a constrained optimization involving Lagrange multipliers~\cite{hofener2012molecular} in which we arrive at the same expression for the embedding potential in Eq.~\ref{embpot}, but formally involving $\rho_{I}$ obtained with coupled cluster theory. However, since in order to obtain coupled cluster densities  it is necessary to solve both $T$ and $\Lambda$ equations, more approximated schemes employing only DFT densities have been put forward~\cite{Gomes2008,Hofener2013,Bouchafra2018} and show very good performance at a much reduced computational cost. In the latter case, the DFT-in-DFT $v_{emb}$ is simply included as an additional one-electron operator to the Fock matrix in the correlated calculation~\cite{Gomes2008}.

\clearpage

\section{Comparison of 2- and 4-component calculations for uranyl tetrachloride}

\begin{figure}[H]
    \centering
    \includegraphics[width=\textwidth,height=\textheight,keepaspectratio]{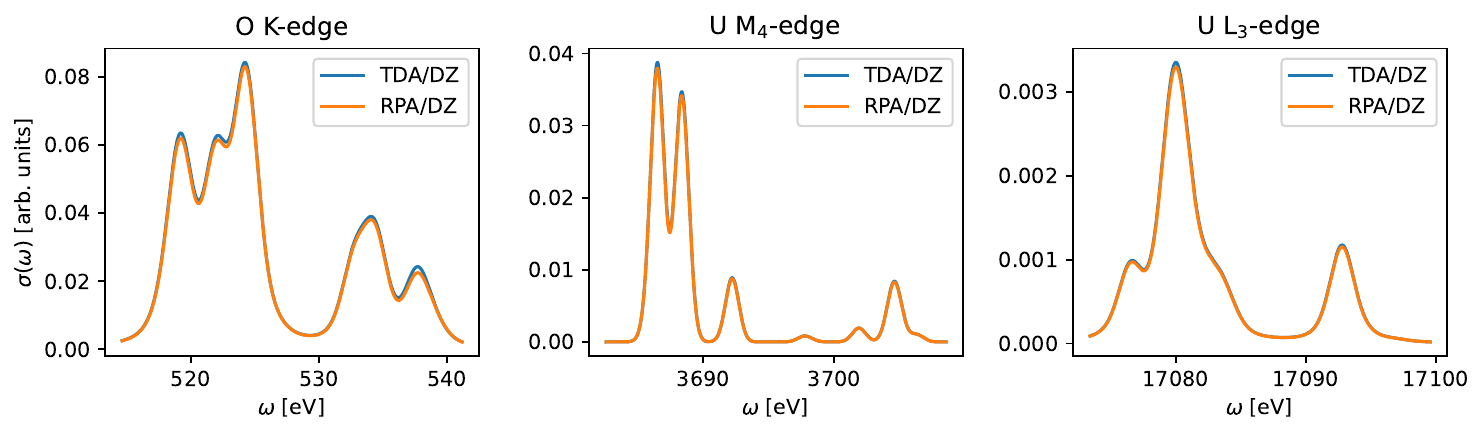}
    \caption{Comparison of 2c-TDA and 2c-TD (RPA) XAS spectra at the edges here investigated using a double-zeta basis set, both with and without employing the Tamm-Dancoff approximation.}
    \label{fig:tda-vs-tddft}
\end{figure}

In Figure~\ref{fig:tda-vs-tddft}, we compare 2-component XAS spectra obtained with TD-DFT with (2c-TDA) and without (2c-TD) invoking the Tamm-Dancoff approximation, and we see that there is no discernible difference.

\begin{figure*}
\centering
\begin{minipage}{\linewidth}
        \centering
        \includegraphics[width=0.70\textwidth,height=\textheight,keepaspectratio]{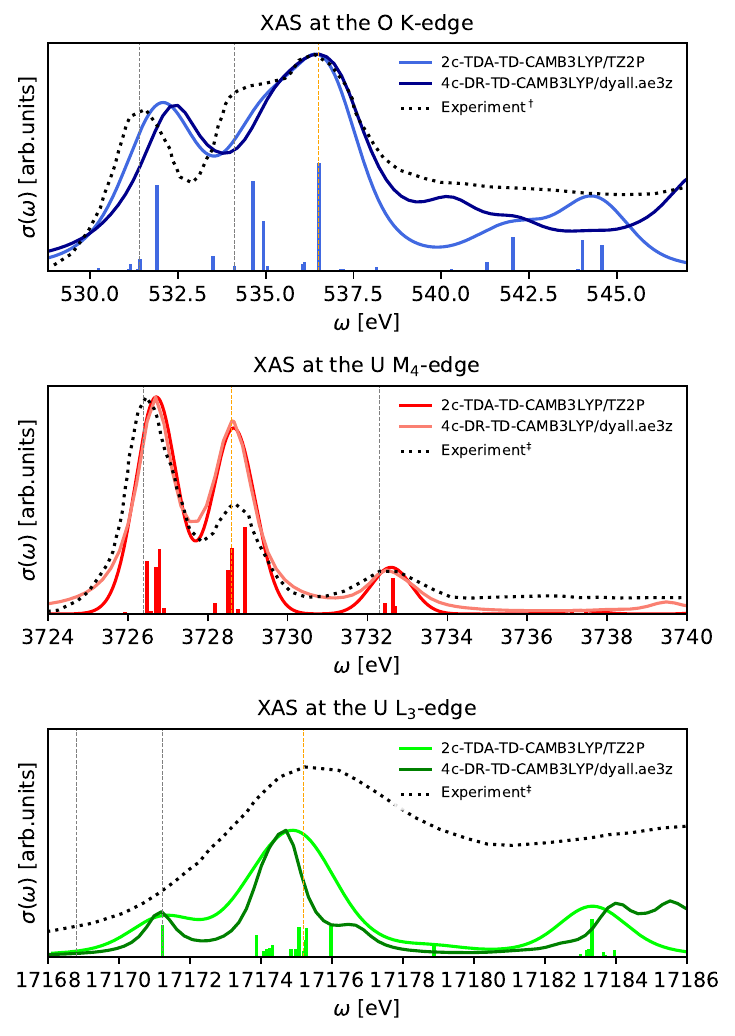}
   \label{fig:ul3o2cl4}
\end{minipage}
    \caption{From top to bottom: 4c-DR and 2c-TDA XAS spectra at the O K-edge, U M$_4$- and L$_3$-edges of UO$_2$Cl$_4^{2-}$. Experimental data at the O K-edge and U M$_4$ and L$_3$- edges of Cs$_2$UO$_2$Cl$_4$ digitized from $^\dagger$\citeauthor{denning2002covalency}\cite{denning2002covalency} and $^\ddag$\citeauthor{vitova2015polarization} \cite{vitova2015polarization}, respectively.}
    \label{fig:4c2cntosuo2cl4}
\end{figure*}

In figure~\ref{fig:4c2cntosuo2cl4} we compare the simulated spectra for the O K-edge, U M$_4$- and L$_3$-edges obtained from 2- and 4-component calculations. It can be seen that overall there are non-negligible but overall small changes in peak positions due to the Hamiltonian (and possibly due to the use of Slater orbitals in ADF) for all edges, though the spectral changes are rather well reproduced--differences here come from the fact that in 2c-TDA one is limited on the number of roots solved for, and on how the stick spectrum obtained with 2c-TDA is subsequently broadened. Further details on the assessment of broadening can be found below.

Except for U L$_3$-edge case, where the experimental broadening is considerably large, the gamma factors was chosen to better fit the experimental spectra, as discussed in the manuscript text.

\pagebreak

\section{Damping parameter for DR-TD-DFT calculations}

\begin{figure}[H]
        \centering
	\includegraphics[width=0.70\textwidth,height=\textheight,keepaspectratio]{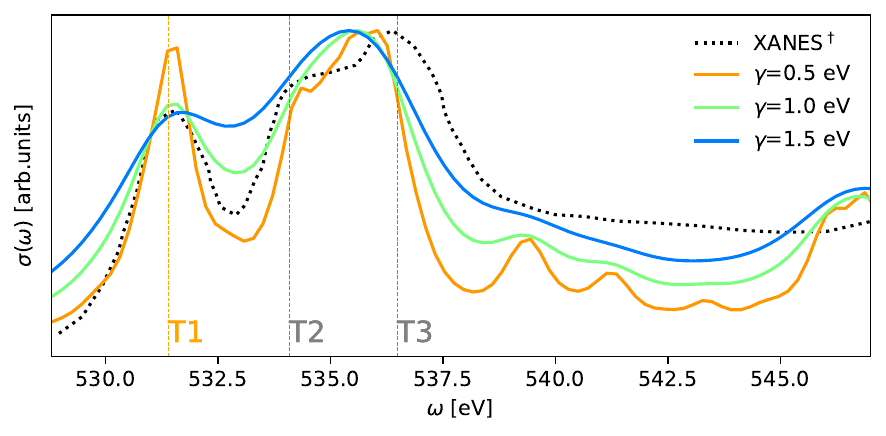}
	\includegraphics[width=0.70\textwidth,height=\textheight,keepaspectratio]{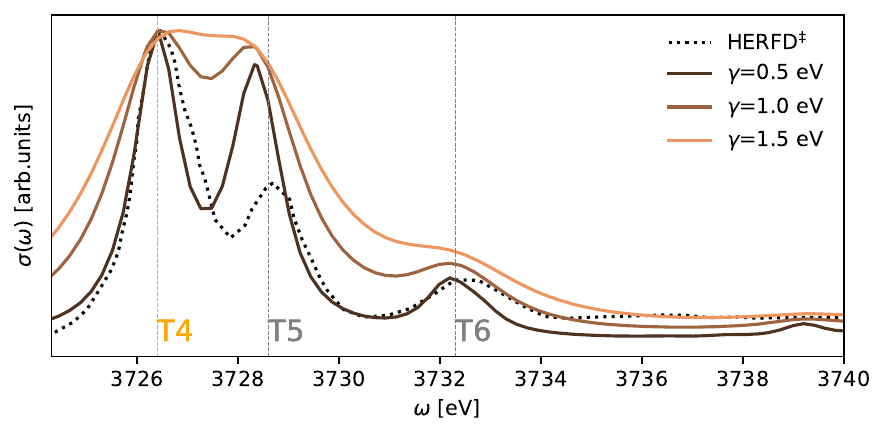}
        \includegraphics[width=0.70\textwidth,height=\textheight,keepaspectratio]{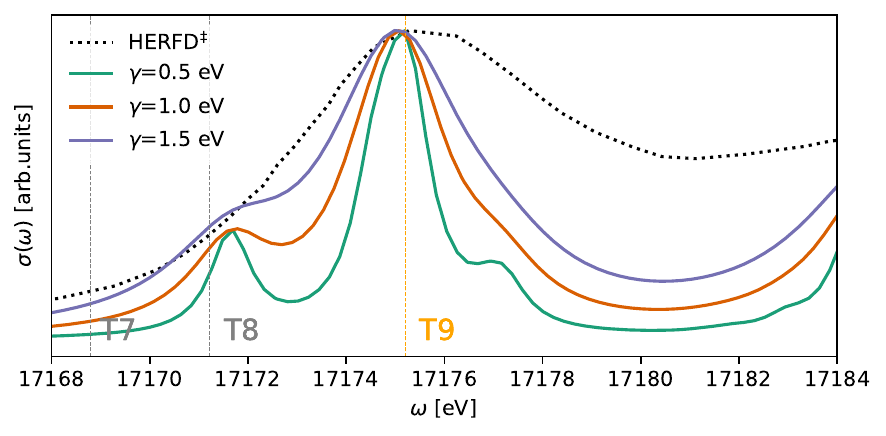}
    \caption{Comparison of 4c-DR XAS spectra at the Oxygen K-edge, Uranium M$_4$- and L$_3$-edge for different damping factors. Shift (depicted as the orange straight line) with respect to the experimental data (Cs$_2$UO$_2$Cl$_4$) from~~\citet{vitova2015polarization}\label{damping-1}}
\end{figure}

\pagebreak

\section{Comparison of broadening functions}

\begin{figure}[H]
	\centering
	\includegraphics[width=0.55\textwidth,height=\textheight,keepaspectratio]{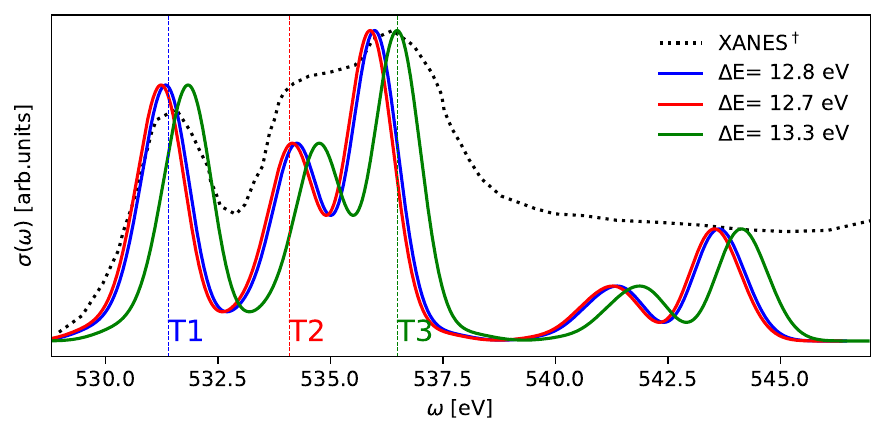}
    \caption*{(a) Gaussian function.}
	\includegraphics[width=0.55\textwidth,height=\textheight,keepaspectratio]{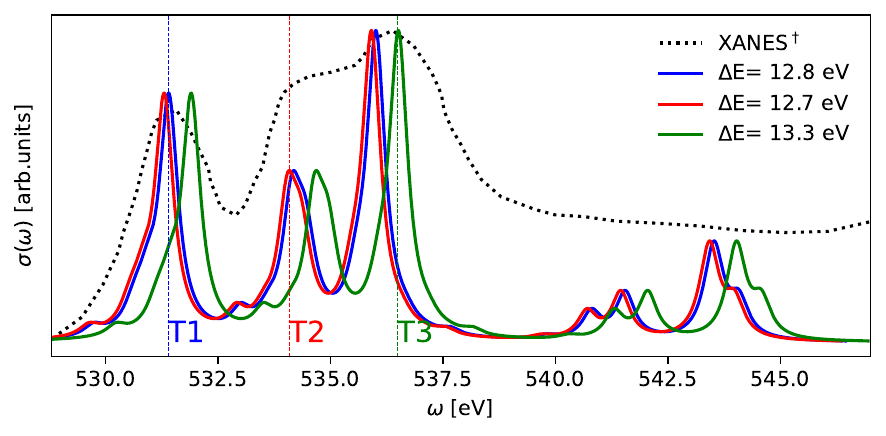}
    \caption*{(b) Lorentzian function.}
	\includegraphics[width=0.55\textwidth,height=\textheight,keepaspectratio]{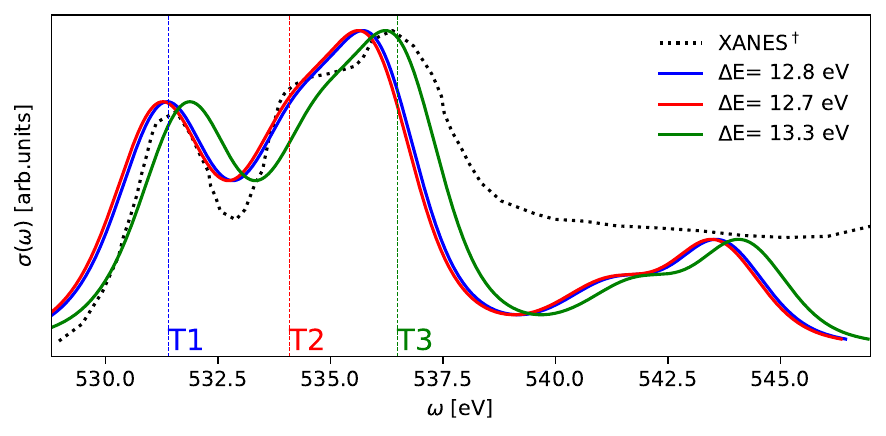}
    \caption*{(c) Voigt profile.}
    \caption{Comparison of 2c-TDA XAS spectra at the oxygen K-edge for different broadening functions ((a) Gaussian; (b) Lorentzian; (c) Voigt profile) and shifts (depicted as the colored straight line) with respect to the experimental data (Cs$_2$UO$_2$Cl$_4$) from~~\citet{denning2002covalency}\label{broadening-1}}
\end{figure}

\begin{figure}
	\centering
	\includegraphics[width=0.55\textwidth,height=\textheight,keepaspectratio]{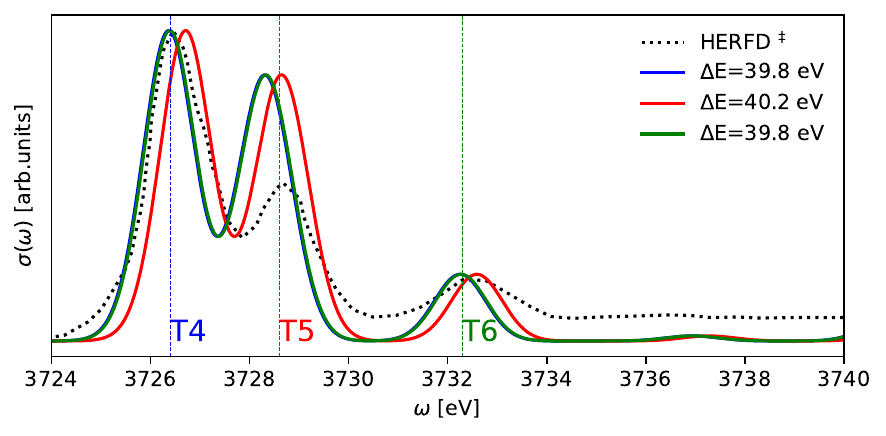}
    \caption*{(a) Gaussian function.}
	\includegraphics[width=0.55\textwidth,height=\textheight,keepaspectratio]{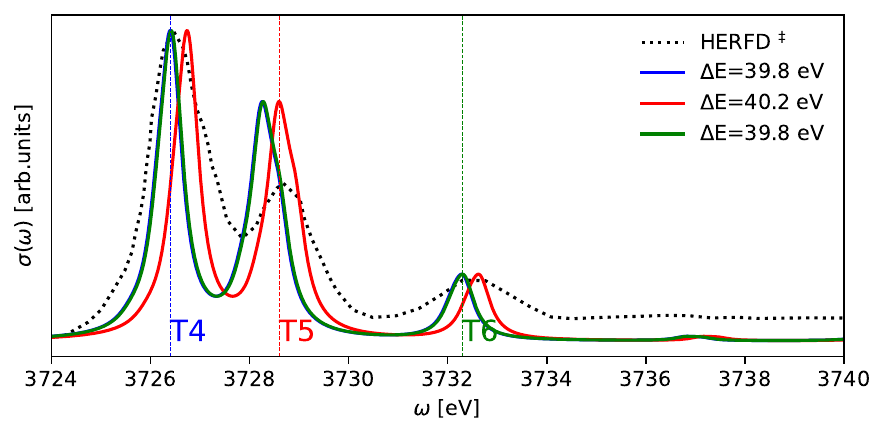}
     \caption*{(b) Lorentzian function.}	
	\includegraphics[width=0.55\textwidth,height=\textheight,keepaspectratio]{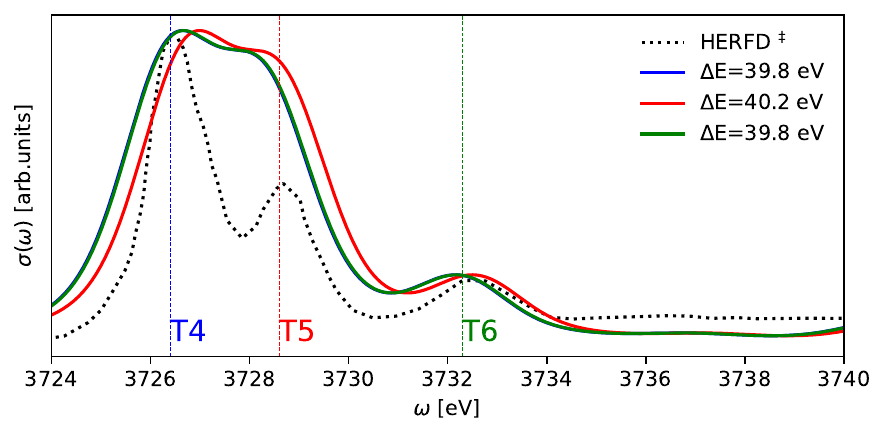}
     \caption*{(c) Voigt profile.}
    \caption{Comparison of 2c-TDA XAS spectra at the uranium M$_4$-edge for different broadening functions ((a) Gaussian; (b) Lorentzian; (c) Voigt profile) and shifts (depicted as the colored straight line) with respect to the experimental data (Cs$_2$UO$_2$Cl$_4$) from~~\citet{vitova2015polarization}\label{broadening-2}}
\end{figure}

\begin{figure}
	\centering
	\includegraphics[width=0.55\textwidth,height=\textheight,keepaspectratio]{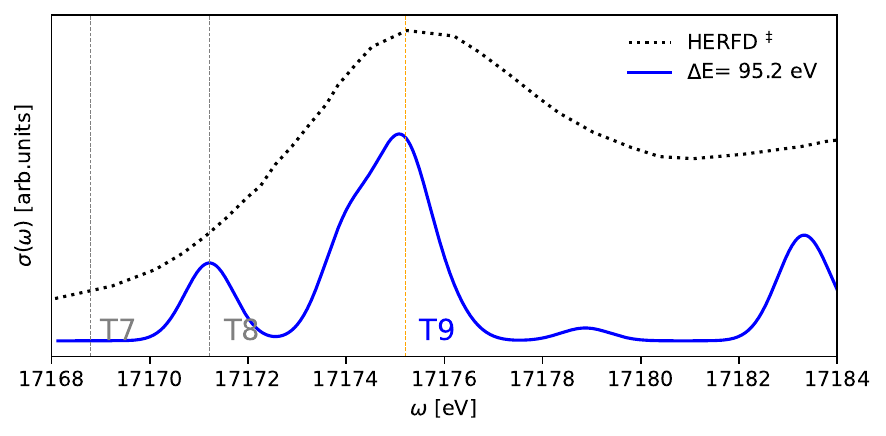}
    \caption*{(a) Gaussian function.}
	\includegraphics[width=0.55\textwidth,height=\textheight,keepaspectratio]{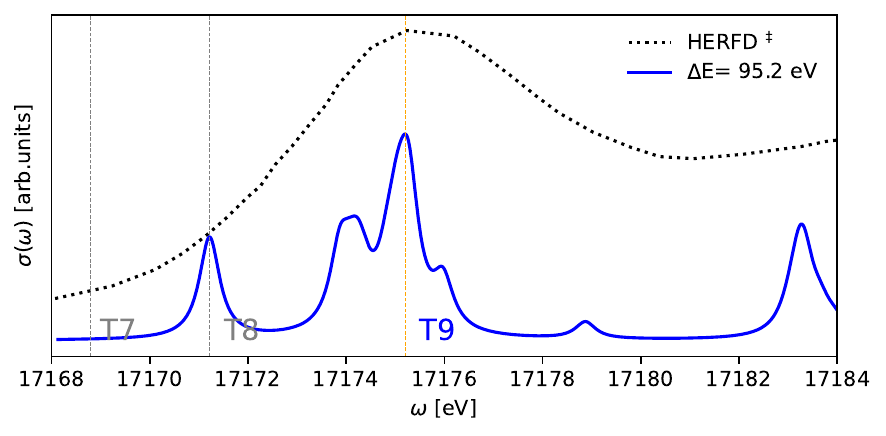}
    \caption*{(b) Lorentzian function.}	
 \includegraphics[width=0.55\textwidth,height=\textheight,keepaspectratio]{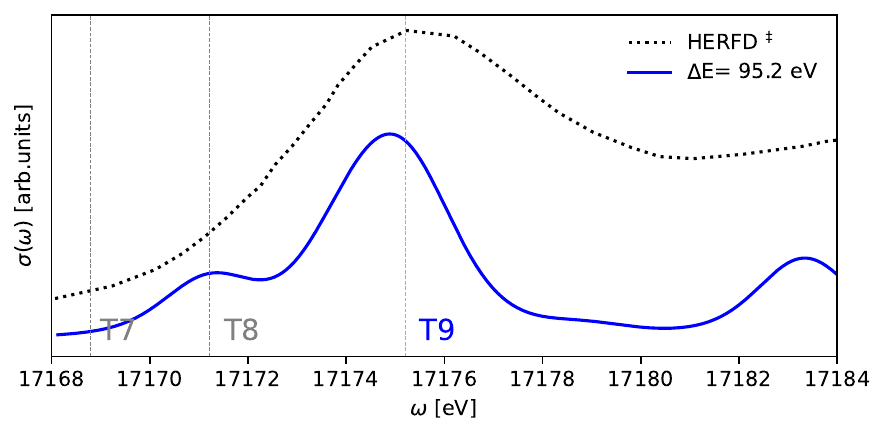}
     \caption*{(c) Voigt profile.}
    \caption{Comparison of 2c-TDA XAS spectra at the uranium L$_3$-edge for different broadening functions ((a) Gaussian; (b) Lorentzian; (c) Voigt profile) with respect to the experimental data (Cs$_2$UO$_2$Cl$_4$) from~~\citet{vitova2015polarization}. Shift to the main line in the experiment, depicted as the orange straight line.\label{broadening-3}}
\end{figure}

\clearpage

\section{Comparison of 4-component calculations for the three models investigated}

In following table we presented the shifts for all calculations performed in this work. In the manuscript text we present the UO$_2$Cl$_4^{2-}$ results.

\begin{table*}[h]
\centering
\caption{Shifts from 2c-TDA-CAMB3LYP and 4-DR-TD-CAMB3LYP XAS spectra with respect to the experimental data (Cs$_2$UO$_2$Cl$_4$) from \citet{denning2007electronic, vitova2015polarization}.}
\begin{tabular}{cccc} 
\hline\hline
                            &              \multicolumn{3}{c}{O K-edge}                               \\ 
\cline{2-4}
                            & UO$_2^{2+}$    & UO$_2^{2+}$ @ Cl$_{4}^{4-}$ & UO$_2$Cl$_4^{2-}$  \\ 
\hline
2c-TDA                      & -            & -                    & 13.5             \\
4c-DR-TD ($\gamma$=1.0 eV) & 13.0           & 13.6                 & 13.8             \\ 
\hline
                            &  \multicolumn{3}{c}{U M$_4$-edge}                                   \\
\cline{2-4}
2c-TDA                      & -            & -                    & 39.8             \\
4c-DR-TD ($\gamma$=0.5 eV) & 36.8         & 36.6                 & 37.2             \\ 
\hline
                            &  \multicolumn{3}{c}{U L$_3$-edge}                           \\
\cline{2-4}
2c-TDA                      & -            & -                    & 95.2             \\
4c-DR-TD ($\gamma$=0.5 eV) & 89.9         & 84.5                 & 86.3             \\
\hline\hline
\end{tabular}
\end{table*}

\clearpage
\pagebreak

\bibliography{refs.bib}